\documentclass[aps,prl,showpacs,revsymb,amsfonts,amssymb,twocolumn,
groupedaddress,footinbib,floatfix]{revtex4-1}

\usepackage{graphicx}
\usepackage{amsfonts}
\usepackage{amssymb}
\usepackage{amsmath}
\usepackage{verbatim}
\usepackage{yfonts}

\usepackage[pdftex,bookmarks,colorlinks,breaklinks]{hyperref}  
\hypersetup{linkcolor=blue,citecolor=red,filecolor=green,urlcolor=green} 

\begin{document}

\title{Campbell response in type II superconductors under strong pinning conditions}

\author{R.\ Willa$^1$}
\author{V.B.\ Geshkenbein$^1$}
\author{R.\ Prozorov$^2$}
\author{G.\ Blatter$^1$}
\affiliation{$^1$Institute for Theoretical Physics, ETH Zurich, 
8093 Zurich, Switzerland} 
\affiliation{$^2$The Ames Laboratory and Department of Physics 
and Astronomy, Iowa State University, Ames, Iowa 50011, USA} 

\date{\today}

\begin{abstract}
Measuring the $ac$ magnetic response of a type II superconductor provides
valuable information on the pinning landscape (pinscape) of the material. We
use strong pinning theory to derive a microscopic expression for the Campbell
length $\lambda_{\rm \scriptscriptstyle C}$, the penetration depth of the $ac$
signal.  We show that $\lambda_{\rm \scriptscriptstyle C}$ is determined by
the jump in the pinning force, in contrast to the critical current $j_c$ which
involves the jump in pinning energy. We demonstrate that the Campbell lengths
generically differ for zero-field-cooled and field-cooled samples and predict
that hysteretic behavior can appear in the latter situation. We compare our
findings with new experimental data and show the potential of this technique
in providing information on the material's pinscape.

\end{abstract}

\pacs{74.25.N-, 
74.25.Op, 
74.25.Wx, 
74.25.Ha  
}

\maketitle


Technologically useful superconductors are of second type and acquire their
desired transport and magnetic properties through vortex pinning, i.e.,
vortices \cite{abrikosov_57} get immobilized by material defects.  The
characterization of the pinning landscape (or pinscape) is of great importance
but presents quite a formidable task.  Measurements of $dc$ transport
properties, either dynamically through the current--voltage characteristic
\cite{kim_64} or statically through magnetization \cite{bean_62}, are standard
techniques used to gain information on the pinscape. Similarly, the $ac$
magnetic response of superconducting samples \cite{campbell_69} provides
insight into the shape of pinning potentials.  Unfortunately, the relation
between the measured penetration depth of the $ac$ signal, the so-called
Campbell length $\lambda_{\rm \scriptscriptstyle C}$, and the parameters of
the pinscape is only known on a phenomenological level.  In this letter, we
present a microscopic derivation of the Campbell length within the framework
of strong pinning theory, thereby providing access to microscopic parameters
of pinning defects and substantially enlarging the scope of applications of
this measurement technique.

Probing superconductors via their $ac$ magnetic response goes back to the
60-ies and culminated in Campbell's work \cite{campbell_69} which provided the
first consistent explanation of the penetration phenomenon (see Refs.\
\cite{further} for further developments): for small $ac$ magnetic-field
amplitudes $h_\mathrm{ac}$ and frequencies $\omega$, vortices oscillate
reversibly within their pinning potentials (described as harmonic wells
$\alpha \, x^2/2$), with the external signal $h_{ac}$ penetrating the sample
on a distance $\lambda_{\rm \scriptscriptstyle C}\propto B/\sqrt{\alpha}$ of
order micrometers. Later work by Lowell \cite{lowell_72} and Campbell
\cite{campbell_78} provided a more quantitative but still phenomenological
understanding within a model pinscape. Here, we make use of the strong pinning
scenario allowing us to perform a quantitative and microscopic analysis of the
$ac$ magnetic response. In particular, we find the dependence of the Campbell
penetration depth $\lambda_{\rm \scriptscriptstyle C}$ on the vortex state,
e.g., the critical (Bean \cite{bean_62}) state with a linear vortex density
gradient supporting the critical current density $j_c$ or a field-cooled state
with a constant induction $B$, and predict the occurrence of new hysteretic
effects. The comparison with recent experiments \cite{prozorov_13} confirms
our predictions.

We consider a geometry with the superconductor occupying the half-space $X>0$,
the magnetic induction $B(X,t) = B_0 + \delta B(X,t)$ directed along $Z$, and
the screening current $j$ flowing along $Y$ (capital and lower case letters
distinguish between macroscopic and microscopic coordinates). The equation of
motion for the macroscopic vortex displacement $U(X,t)$ reads
\begin{align}\label{eq:eom}
  \eta \partial_t U = F_{\rm \scriptscriptstyle L} (j,U) + F_\mathrm{pin} (X,U),
\end{align}
with the Lorentz force $F_{\rm \scriptscriptstyle L}$ balanced by dissipative
and pinning forces ($\eta$ denotes the viscosity \cite{bardeen_65}). The
displacement $U(X,t)$ relates to the induction via $\delta
B(X,t) = - B_0\, \partial_{\rm \scriptscriptstyle X} U(X,t)$ and is driven at
the surface $X=0$ by the small external field $h_{ac} \ll B_0$, $\delta B(0,t)
= h_\mathrm{ac} e^{-i\omega t}$.  The Lorentz force $F_{\rm \scriptscriptstyle
L} = (j_0+\delta j) B/c$ involves an $ac$ component $\delta j = - c
\partial_{\rm \scriptscriptstyle X} \delta B/4\pi$ and writing the pinning
force $F_\mathrm{pin} = F_0 +\delta F_\mathrm{pin}$, with $F_0$ the force
density in the initial vortex state balancing the $dc$ Lorentz force $j_0
B_0/c$, we obtain the dynamical equation
\begin{align}\label{eq:eomU}
   \eta \partial_t U - (B_0^2/4\pi) \partial^{2}_{\scriptscriptstyle X} U 
   - \delta F_\mathrm{pin} (U) = 0.
\end{align}
Following \cite{campbell_69}, one assumes small oscillations of the vortices
near the potential minima. This motivates the phenomenological Ansatz $\delta
F_\mathrm{pin}(U) = - \alpha U$ for the pinning force density. Solving 
(\ref{eq:eomU}) for the displacement field,
\begin{align}\label{eq:lC}
   U(X,t) &=  \lambda_{\rm \scriptscriptstyle C}(h_{ac}/B_0) 
   e^{-X/\lambda_{\rm \scriptscriptstyle C}} e^{-i \omega t}\\ \label{eq:lC_gen}
   \mathrm{with~~~} \lambda_{\rm \scriptscriptstyle C}^2(\omega) 
          &= B_0^2/4\pi(\alpha - i \omega \eta),
\end{align}
results in the Campbell length $\lambda_{\rm \scriptscriptstyle C} =
\lambda_{\rm \scriptscriptstyle C} (\omega = 0) = (B_0^2/4\pi \alpha)^{1/2}$
at low frequencies.

Here, our goal is to derive an expression for $\delta F_\mathrm{pin}$ starting
from a microscopic perspective. This can be done within the framework of
strong pinning theory which goes back to work of Labusch \cite{labusch_69} and
Larkin and Ovchinnikov \cite{larkin_79}, with recent further studies on the
critical currents in strong and weak pinning scenaria \cite{blatter_04},
numerical simulations of vortex motion \cite{koshelev_11}, and the
current--voltage characteristic \cite{thomann_12}; note that the qualitative
framework of weak collective pinning theory \cite{larkin_79} is not sufficient
to develop a quantitative understanding of $\lambda_{\rm \scriptscriptstyle
C}$.

\begin{figure}[tb]
\includegraphics[width=8.0cm]{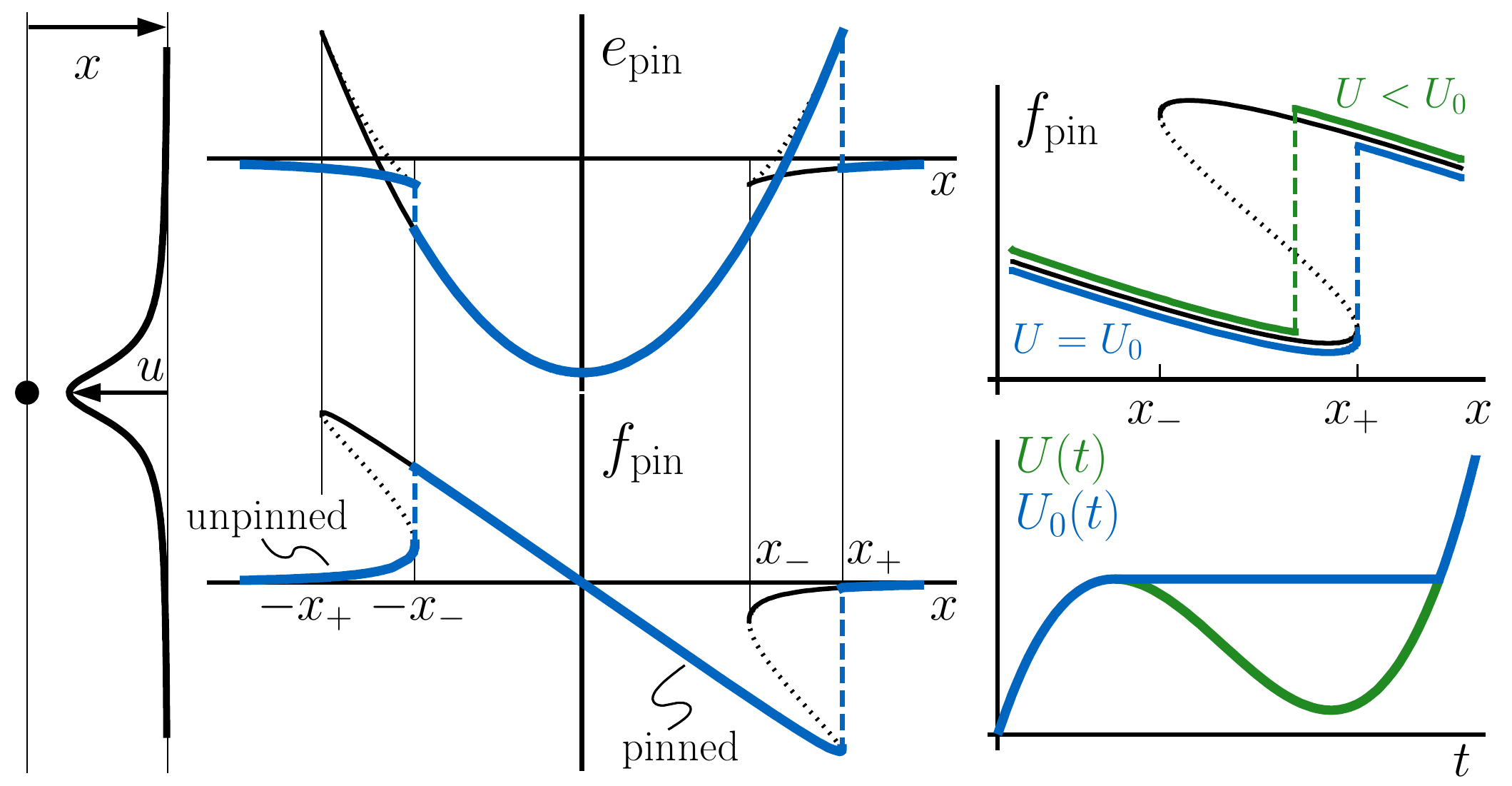}
\caption{Pinning energy $e_\mathrm{pin}$ and force $f_\mathrm{pin}$ in a
strong pinning situation for a Lorentzian potential. The bistable solutions
near the defect describe pinned and unpinned branches.  Thick lines (blue)
mark occupied branches in the Bean state, dotted lines are unstable solutions,
dashed lines are the jumps making up for $\Delta e_\mathrm{pin}$ and $\Delta
f_\mathrm{pin}$.  Left: strong pinning situation for a representative vortex.
The microscopic displacement $u(x) = f_\mathrm{pin}(x)/\bar{C}$ has the same
shape as $f_\mathrm{pin}(x)$. Right: change in branch occupation when the
vortex system moves by $U$.}
\label{fig:ZFC}
\end{figure}

Consider a representative vortex within the flux-lattice driven along $x$ on a
trajectory described through the asymptotic coordinate ${\bf r}_\infty =
(x,b)$ at large $|z|$; the distance $b$ along $y$ is the impact parameter with
respect to a defect at the origin.  Within the strong pinning context, defects
act individually, generating a pinning potential $e_p(r,z)$.  Considering a
trajectory with maximal pinning, i.e., $b = 0$ and including the deformation
energy of the vortex, its total energy as a function of $x$ takes the
form (we assume a point-like defect with $e_p (x,z) = e_p(x) \delta(z)$
\cite{blatter_04})
\begin{align}\label{eq:upin}
   e_\mathrm{pin}(x) = \frac{1}{2} \bar{C} u(x)^{2} + e_p[x+u(x)],
\end{align}
with $u(x)$ the microscopic displacement field in the plane $z=0$, see Fig.\
\ref{fig:ZFC}, and $\bar{C}$ the effective elasticity of the vortex embedded
within the lattice,
\begin{align}\label{eq:cbar-def}
  \bar{C}^{-1} = \frac{1}{2} \int \frac{d^{3}k}{(2\pi)^{3}} 
  \frac{1}{c_{66}(k_{x}^{2}+k_{y}^{2}) + c_{44}({\bf k})k_{z}^{2}}.
\end{align}
Here, $c_{66}$ and $c_{44}({\bf k})$ denote shear and dispersive tilt
moduli and proper integration in (\ref{eq:cbar-def}) provides the 
result $\bar{C} \sim (a_0^2/\lambda) \sqrt{c_{66} c_{44}(0)}$ with $a_0^{-2} =
B_0/\Phi_0$ the vortex density ($\Phi_0 = hc/2e$ is the flux unit and
$\lambda$ the London penetration depth). Minimization of (\ref{eq:upin}) with respect
to $u$ (at fixed $x$) generates the self-consistency condition
\begin{align}\label{eq:Labusch-eq}
   \bar{C} u(x) = f_p[x+u(x)]
\end{align}
for the displacement field $u(x)$, where $f_p(x) = -e_p^\prime(x)$ is the bare
force profile of the pinning defect, the prime denoting derivative with
respect to $x$. The maximal slope in $f_p^\prime$ (realized at $x_m$) defines
the regime of strong pinning \cite{labusch_69}: for $\kappa \equiv [f_p^\prime
(x_m)]/\bar{C} > 1$, the condition (\ref{eq:Labusch-eq}) generates two stable
solutions for the displacement field $u(x)$, a pinned and an unpinned branch,
see Fig.\ \ref{fig:ZFC}.  The condition $\kappa = 1$ is the famous Labusch
criterion \cite{labusch_69} separating strong pins with $\kappa > 1$ from weak
pins when $\kappa < 1$.

Assuming a homogeneous random distribution of defects with small density
$n_p$, the macroscopic pinning force density $F_\mathrm{pin}$ derives from
averaging the pinning forces $f_p[x+u_\mathrm{o}(x)]$ over all
positions $|x| < a_0/2$ within a lattice period, with $u_\mathrm{o}$ denoting
the branch that is occupied with vortices. This occupation depends on the
state preparation, e.g., for a Bean state with vortices driven
along $x$, the occupation of the pinned branch extends over the
interval $[-x_-,x_+]$, see Fig.\ \ref{fig:ZFC}, such as to produce the maximal
force $F_\mathrm{pin} = F_c$,
\begin{align}\label{eq:dc-pinning}
   F_c = n_p \langle f_\mathrm{pin} \rangle = n_p \frac{t_\perp}{a_0}  
   \int_{a_0} \frac{dx}{a_0} f_\mathrm{pin}(x)|_\mathrm{o},
\end{align}
where $f_\mathrm{pin}(x) \equiv f_p[x+u(x)]$ and $|_\mathrm{o}$ refers to the
occupied branch $u_\mathrm{o}(x)$ (we assume maximal pinning for all
trajectories with $2|b| < t_\perp \simeq \xi$, $\xi$ the coherence length).
Making use of the relation $f_\mathrm{pin}(x) = -d e_\mathrm{pin}(x)/dx$, we
arrive at a simple expression for the critical current density $j_c = (c/B)
F_c$,
\begin{align}
   j_c = \frac{c}{B} n_p \frac{t_\perp}{a_0^2} \int_{a_0} \!\!\!\!
         dx \, [-d e_\mathrm{pin}(x)/dx|_\mathrm{o}]
       = \frac{c n_p t_\perp}{\Phi_0} \Delta e_\mathrm{pin}, \label{eq:jc}
\end{align}
where $\Delta e_\mathrm{pin}$ is the sum of jumps at $-x_-$ and $x_+$ in
$e_\mathrm{pin}(x)$ where the occupation changes between unoccupied and
occupied branches \cite{labusch_69,larkin_79}, see Fig.\ \ref{fig:ZFC}.

Equipped with this microscopic understanding of pinned vortex matter in the
Bean state, we return to the problem of $ac$ magnetic response.  Within strong
pinning, we can follow the changes in the occupation of pinned and unpinned
branches as vortices are driven by the $ac$-magnetic field and calculate the
time dependent and inhomogeneous change in the pinning force $\delta
F_\mathrm{pin} [U(x,t)]$.  A macroscopic shift $U > 0$ pushes vortices in the
direction of the Lorentz force; vortices at $-x_-$ and $x_+$ jump to pinned
and unpinned branches, respectively, leaving the branch occupation unchanged,
hence $\delta F_\mathrm{pin}(U \! > \! 0) = 0$. Otherwise, a negative
displacement $U \! < \!0$ shifts the boundaries between occupied and
unoccupied states to the left, see Fig.\ \ref{fig:ZFC}.  This results in a
change of the macroscopic restoring force
\begin{align}\label{eq:dFp_def}
   \delta F_\mathrm{pin}(U \! < \! 0)
   &= n_p\frac{t_\perp}{a_0^2}\! \int_{a_0} \!\!\!\! dx 
   [f_\mathrm{pin}(x)|_{\mathrm{o},U}\! -\! f_\mathrm{pin}(x)|_{\mathrm{o},0}],
\end{align}
where the index $|_{\mathrm{o},U}$ refers to the occupation where vortices
have been shifted by $U$. Expanding the integrand for small $U$, we arrive at
the expression
\begin{align}\label{eq:FnU}
   \delta F_\mathrm{pin}(U<0)
   &= n_p \frac{t_\perp}{a_0^2} \int_{a_0} \!\!\!
         dx \, d f_\mathrm{pin}(x)/dx\big|_\mathrm{o}\, U,
\end{align}
resulting in the strong pinning result for $\delta F_\mathrm{pin}$,
\begin{align}
   \delta F_\mathrm{pin}(U) &\approx -n_p (t_\perp/a_0^2) \Delta f_\mathrm{pin} 
   \, \Theta(-U)\, U, 
  \label{eq:dFp}
\end{align}
with $\Delta f_\mathrm{pin}$ the sum of jumps in the function
$f_\mathrm{pin}$.

Inserting this result into Eq.\ (\ref{eq:eomU}) generates a complex vortex
dynamics as flux enters the sample in a sequence of diffusive pulses until the
field is raised to $B_0 + h_{ac}$, see \cite{willa_15} for a detailed
description of this initialization process. After saturating the sample at
this higher field level, the displacement $U(X,t)$ assumes the form
\begin{align}\label{eq:U_B}
   U(X,t) = U_0(X) - \lambda_{\rm\scriptscriptstyle C} (h_{ac}/B_0) 
   e^{-X/\lambda_{\rm \scriptscriptstyle C}} [1-e^{-i \omega t}],
\end{align}
with $U_0(X) = (h_{ac}/B_0)(L-X)$ generating the shift in field $B_0 \to
B_0(1-\partial_{\rm \scriptscriptstyle X}U_0) = B_0 + h_{ac}$ and $L$ the
penetration depth of the Bean profile \cite{remark}.  The second term accounts
for the penetration of the external field with respect to the {\it new} Bean
state, $\delta B(X,t) = h_{ac}e^{-X/\lambda_{\rm \scriptscriptstyle C}}
(1-e^{-i\omega t})$. The Campbell penetration depth can be expressed by
the microscopic parameters, the average curvature $d^{\, 2}
e_\mathrm{pin}(x)/dx^2$, of the pinscape,
\begin{align}
   \frac{B_0^2}{4 \pi \lambda_{\rm \scriptscriptstyle C}^2} 
   \! =\!  -\frac{n_p t_\perp}{a_0^2}\!\! \int_{a_0} \!\!\!\!
         dx \, d^{\, 2} e_\mathrm{pin}(x)/dx^2\big|_\mathrm{o} \!\!
   = \frac{n_p t_\perp}{a_0^2} \Delta f_\mathrm{pin}.
   \label{eq:lc}
\end{align}
Making use of the estimates $\Delta f_\mathrm{pin} \sim f_p$, $t_\perp \sim
\xi$, and $\kappa \sim f_p/\xi \bar C$, we find that $\lambda_{\rm
\scriptscriptstyle C}^2 \sim \lambda^2/(\kappa n_p a_0 \xi^2) > \lambda^2$
with $\kappa \, n_p a_0 \xi^2\ll 1$ the small parameter defining the
three-dimensional strong pinning regime \cite{blatter_04}.  Comparing the
results for $j_c$ and $\lambda_{\rm \scriptscriptstyle C}^{-2}$, Eqs.\
(\ref{eq:jc}) and (\ref{eq:lc}), we observe that these two quantities address
different properties of the pinscape, the jumps in pinning energy and force,
respectively. As a consequence, the simple scaling $j_c \sim c \alpha \xi/B
\sim (c/4\pi) \xi B/\lambda_{\rm \scriptscriptstyle C}^2$ previously
conjectured on the basis of the phenomenological result (\ref{eq:lC_gen})
turns out incorrect and has to be replaced by $j_c \sim (c/4\pi) \kappa \xi
B/\lambda_{\rm \scriptscriptstyle C}^2 \propto [\Delta f_\mathrm{pin}]^2$.
Hence, care must be taken when translating measured data on $\lambda_{\rm
\scriptscriptstyle C}$ into predictions for $j_c$ \cite{prozorov_13}.
\begin{figure}[tb]
\includegraphics[width=8cm]{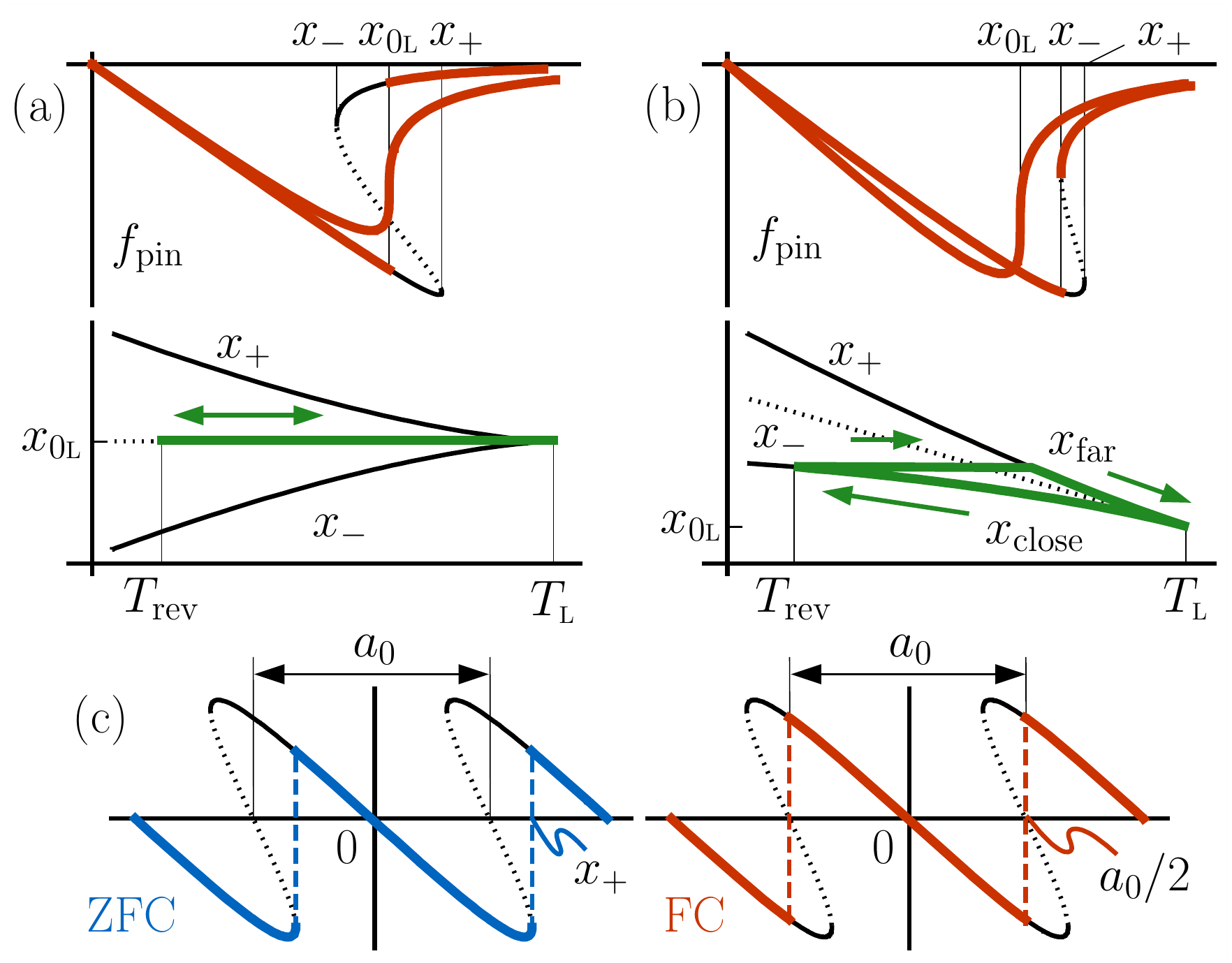}
\caption{Evolution of the pinning force $f_\mathrm{pin}$ crossing over from
weak to strong pinning. The jump in the occupation between pinned and unpinned
branches first appears at $x_{0\rm \scriptscriptstyle L}$ and remains there if
the branch edges at $x_\pm$ move away in opposite directions with decreasing
temperature, $x_- < x_{0\rm \scriptscriptstyle L} < x_+$, see (a).  If
$x_{0\rm \scriptscriptstyle L} < x_- < x_+$, see (b), the jump is pinned to
$x_-$ and hysteretic effects show up upon thermal cycling. (c) Pinscape
$f_\mathrm{pin}(x)$ at high magnetic fields involving only pinned and unstable
branches. The relevant jumps are located at $x_+$ for the zero-field-cooled
sample (left) and at $a_0/2$ for the field-cooled situation (right).}
\label{fig:FC}
\end{figure}

Next, we turn to the field-cooled state with $j_0 = 0$ and $F_0=0$.  Following
(\ref{eq:lc}), the determination of the jumps in the, now symmetric,
occupation of $f_\mathrm{pin}$ is the central task in the calculation of
$\lambda_{\rm \scriptscriptstyle C}$.  Assuming defects in the form of
metallic or insulating inclusions, one can show \cite{willa_15} that pinning
turns on smoothly upon crossing the $H_{c2}(T)$ line.  Hence, the vortex
system changes from weak to strong pinning upon decreasing the temperature $T$
below the Labusch temperature $T_{\rm \scriptscriptstyle L}$ defined through
$\kappa(T_{\rm \scriptscriptstyle L}) = f_p^\prime (x_m)/\bar{C}|_{T_{\rm
\scriptscriptstyle L}} = 1$.  At $T_{\rm \scriptscriptstyle L}$, the pinning
force $f_\mathrm{pin}(x)$ develops an infinite slope at $x_{0\rm
\scriptscriptstyle L}$, $[d f_\mathrm{pin} / dx|_{x_{0\rm \scriptscriptstyle
L}}]_{T_{\rm \scriptscriptstyle L}} = \infty$.  Lowering the temperature below
$T_{\rm \scriptscriptstyle L}$, the function $f_\mathrm{pin}(x)$ develops two
branches, pinned and unpinned ones, which start and end at the boundaries $\pm
x_+$ and $\pm x_-$ close to $\pm x_{0\rm \scriptscriptstyle L}$.  In order to
decide upon the branch occupation below $T_{\rm \scriptscriptstyle L}$, we
have to determine the relative arrangement of the positions $x_{0\rm
\scriptscriptstyle L}$ and $x_\pm$. We distinguish three cases, of which (a)
is the simplest one, see Fig.\ \ref{fig:FC}(a), with $x_\pm$ moving away from
$x_{0\rm \scriptscriptstyle L}$ in different directions.  In this case, the
branch occupation jumps between pinned and unpinned at $\pm x_{0\rm
\scriptscriptstyle L}$ and a small $ac$ field produces a small reoccupation
around these points; the relevant jumps in $f_\mathrm{pin}$ thus appear at
$\pm x_{0\rm \scriptscriptstyle L}$, with $\Delta f_\mathrm{pin} = 2 \Delta
f_\mathrm{pin}|_{x_{0\rm \scriptscriptstyle L}}$ entering the expression for
the field-cooled Campbell length (\ref{eq:lc}). Case (b) shown in Fig.\
\ref{fig:FC}(b) describes the situation where both branches grow beyond
$x_{0\rm \scriptscriptstyle L}$ with decreasing temperature, $x_{0\rm
\scriptscriptstyle L} < x_- < x_+$. Then, vortices between $x_{0 \rm
\scriptscriptstyle L}$ and $x_-$ jump to the pinned branch and the relevant
jump in the occupation is pinned to $x_-$.  Accordingly, the jump in the
pinning force entering $\lambda_{\rm \scriptscriptstyle C}$ is given by $2
\Delta f_\mathrm{pin}|_{x_-}$. Finally, case (b') involves a shrinking of the
branches with respect to $x_{0\rm \scriptscriptstyle L}$, i.e., $x_- < x_+ <
x_{0\rm \scriptscriptstyle L}$, and the jump in occupation is pinned to $x_+$,
$\Delta f_\mathrm{pin} = 2 \Delta f_\mathrm{pin}|_{x_+}$. As a result, the
Campbell length $\lambda_{\rm \scriptscriptstyle C}$ may differ for the
zero-field-cooled (Bean type) and field-cooled vortex states in various
respects, depending on the case at hand.

Quantitative analytic results can be obtained at temperatures below but close
to $T_{\rm \scriptscriptstyle L}$ where $\kappa \gtrsim 1$.  Expanding the bare
pinning force $f_p(x)$ around $x_m$ (where $f_p^{\prime\prime}$ vanishes),
$f_p(x) \approx f_p(x_m) + f_p^\prime|_{x_m} (x-x_m) - \gamma (x-x_m)^3/3$
with $2\gamma = -f_p^{\prime\prime\prime}|_{x_m} >0$, we obtain the result
\begin{align}\label{eq:xpm}
   x_\pm = x_0 \pm
   \frac{2}{3}\sqrt{\frac{\bar{C}}{\gamma}} (\kappa-1)^{3/2},
\end{align}
with $x_0 = x_m-f_p(x_m)/\bar{C} > x_m$ the generalization of $x_{0\rm
\scriptscriptstyle L}$ to temperatures below $T_{\rm \scriptscriptstyle L}$,
$x_0(T_{\rm \scriptscriptstyle L}) = x_{0\rm \scriptscriptstyle L}$.  The
jumps at $\pm x_\pm$ then are equal and smaller than the jumps at $\pm x_{0\rm
\scriptscriptstyle L}$. For case (a), this results in different (by $\approx 7
\%$) Campbell lengths $\lambda_{\rm \scriptscriptstyle C}|_{\rm
\scriptscriptstyle FC} < \lambda_{\rm \scriptscriptstyle C}|_{\rm
\scriptscriptstyle ZFC}$, while for the cases (b) and (b') the two lengths are
equal.  For large $\kappa \gg 1$, the three jumps are all different, resulting
in different Campbell lengths with $\lambda_{\rm \scriptscriptstyle C}|_{\rm
\scriptscriptstyle FC^+} < \lambda_{\rm \scriptscriptstyle C}|_{\rm
\scriptscriptstyle ZFC} < \lambda_{\rm \scriptscriptstyle C}|_{\rm
\scriptscriptstyle FC^-}$, where $\pm$ refer to the scenaria involving the
large and small jumps at $x_\pm$.

Which of the above scenaria is realized in a specific case depends on the
temperature dependence of elastic and pinning forces. Close to $T_{\rm
\scriptscriptstyle L}$, the behavior of $x_\pm$ is dominated by $x_0 \sim
x_{0\rm \scriptscriptstyle L} + a \tau_{\rm \scriptscriptstyle L}$ with
$\tau_{\rm \scriptscriptstyle L} = 1-T/T_{\rm \scriptscriptstyle L}$ and the
sign of the prefactor $a$ deciding upon which case (b) or (b') is realized. On
the other hand, for larger $\tau_{\rm \scriptscriptstyle L}$ the second term
in (\ref{eq:xpm}), $\propto (\kappa-1)^{3/2}\propto \tau_{\rm
\scriptscriptstyle L}^{3/2}$, becomes dominant and case (a) is realized.

Furthermore, hysteretic behavior of $\lambda_{\rm \scriptscriptstyle C}$
appears in cases (b) and (b') when first cooling and subsequently reheating
the sample (from $T_\mathrm{min}$). Indeed, when both branches increase or
decrease below $x_{0\rm \scriptscriptstyle L}$ upon cooling, the relevant jump
appears at the branch edge $x_\mathrm{close}$ that is closer to $x_{0\rm
\scriptscriptstyle L}$. On reheating, the jump first remains pinned to
$x_\mathrm{close} (T_\mathrm{min})$ until the other edge $x_\mathrm{far}$
further away from $x_{0\rm \scriptscriptstyle L}$ is hit, whereupon the jump
follows the position $x_\mathrm{far}(T)$, see Fig.\ \ref{fig:FC}(b).
Otherwise, in case (a) or when $x_\mathrm{close}$ goes through an extremum, no
hysteresis appears upon thermal cycling as long as the jump in
$f_\mathrm{pin}$ is realized \cite{jump} away from the branch edges at $\pm
x_\pm$.

Next, we briefly discuss the situation at high fields when the pinned branch
extends beyond the vortex separation $a_0$, $x_+ > a_0/2$. Close to $H_{c_2}$,
the bare pinning force is well approximated by the lowest harmonic, $f_p(x)
\approx f_0 \sin(2\pi x/a_0)$; the competition with elastic forces then
produces the multi-valued function $f_\mathrm{pin}(x)$ shown in Fig.\
\ref{fig:FC}(c). In this situation, the branch edges at $\pm x_-$ have
vanished and only the pinned branches between $\pm x_+$ survive.  For the Bean
state, the jump in force ($\Delta f_\mathrm{pin}|_{x_+}$) determining
$\lambda_{\rm \scriptscriptstyle C}$ is located at $x_+$. For the field-cooled
state, the (slightly larger) jump in force is located at $a_0/2$ instead,
hence $\lambda_{\rm \scriptscriptstyle C}|_{\rm \scriptscriptstyle FC}
\lesssim \lambda_{\rm \scriptscriptstyle C}|_{\rm \scriptscriptstyle ZFC}$; no
hysteresis is expected in this regime.  Upon decreasing the field, additional
harmonics become relevant in the description of $f_p(x)$ and its maximal slope
at $x_m$ moves away from $a_0/2$, i.e., $x_m < a_0/2$. As $x_{0\rm
\scriptscriptstyle L}$ also decreases below $a_0/2$ an unpinned branch starts
developing and we cross over to the low-field domain involving both the pinned
and unpinned branches. Note that neither of these regimes is small but rather
occupy similar size regions within the $H$-$T$ phase diagram.

\begin{figure}[tb] 
\includegraphics[width=8.5cm]{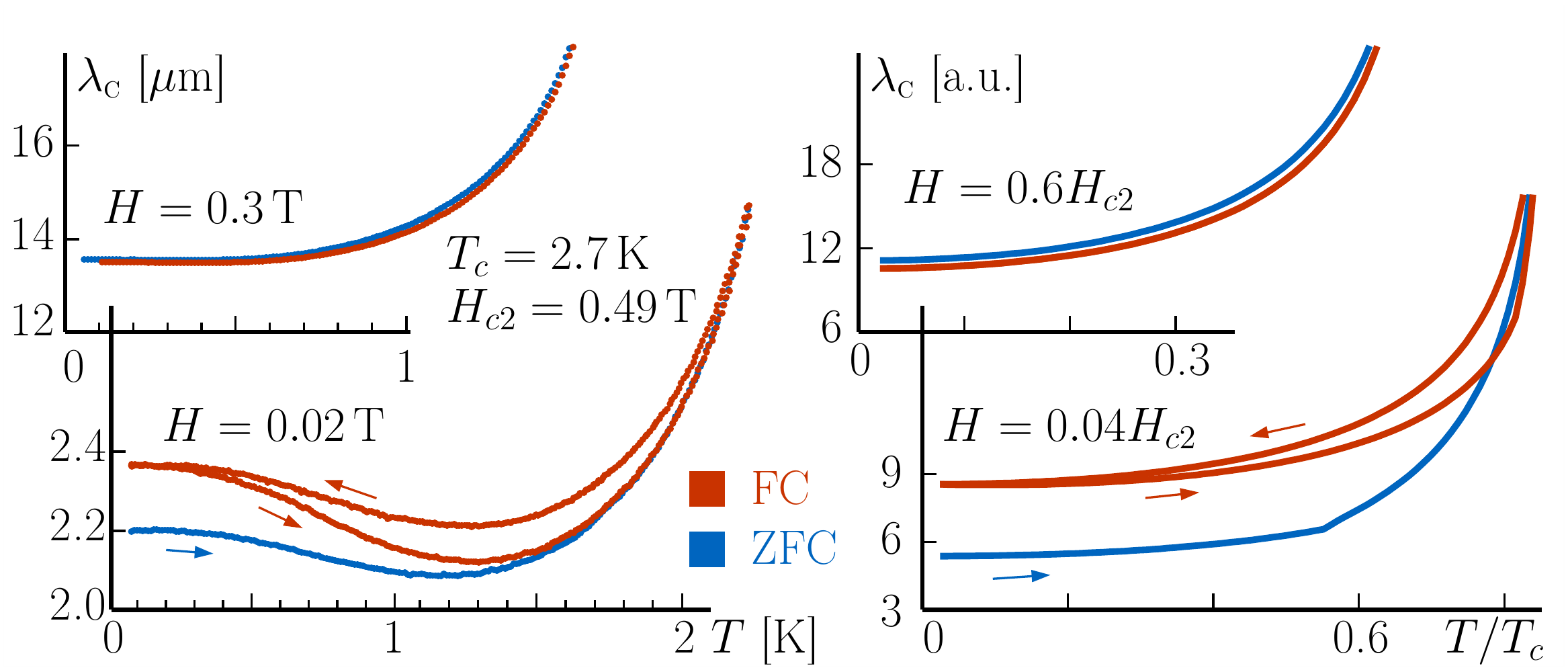}
\caption{Experimental (left) and theoretical (right) traces of the Campbell
length $\lambda_{\rm\scriptscriptstyle C}(T)$ for zero-field-cooled (blue) and
(hysteretic) field-cooled (red) states at low (main panels) and high (inserted
panels) magnetic fields.}
\label{fig:exp}
\end{figure}

In Fig.\ \ref{fig:exp} we compare our main new findings, the dependence of
$\lambda_{\rm \scriptscriptstyle C}$ on the vortex state and the appearance of
hysteretic effects, with measurements on a single crystal superconductor
SrPd$_2$Ge$_2$ (isostructural to the Fe- and Ni-pnictides) using a
tunnel-diode oscillator technique, see Fig.\ 4(a) of Ref.\ \cite{prozorov_13}
(shown are magnified traces at $0.02$ T and $0.3$ T). A small $ac$ excitation
field $h_{ac} \approx 20$ mOe is superimposed on the $dc$ field ensuring
linearity of the response, see Ref.\ \cite{prozorov_03} for experimental
details.  Theoretical results for the Campbell lengths are found by solving
Eq.\ (\ref{eq:Labusch-eq}) and extracting the relevant jumps $\Delta
f_\mathrm{pin}$, assuming a pinning model based on insulating inclusions
\cite{willa_15} (we use standard Ginzburg-Landau scaling).  All features, the
dependence of $\lambda_{\rm \scriptscriptstyle C}$ on the state preparation,
the appearance of hysteresis upon thermal cycling, as well as the reversal
from $\lambda_{\rm \scriptscriptstyle C}|_{\rm \scriptscriptstyle ZFC} <
\lambda_{\rm \scriptscriptstyle C}|_{\rm \scriptscriptstyle FC^-}$ at low
fields to $\lambda_{\rm \scriptscriptstyle C}|_{\rm \scriptscriptstyle FC} <
\lambda_{\rm \scriptscriptstyle C}|_{\rm \scriptscriptstyle ZFC}$ at high
fields, are visible in the experiment and captured by the model; note that
other pinning models based on metallic inclusions or $\delta T_c$-, $\delta
\ell$-pinning \cite{blatter_94} ($\ell$ the mean free path) produce different
behavior.

In conclusion, making use of strong pinning theory, we have presented a
microscopic and quantitative expression for the Campbell length $\lambda_{\rm
\scriptscriptstyle C}$ that captures specific properties of the pinscape. Our
theory predicts the dependence of $\lambda_{\rm \scriptscriptstyle C}$ on the
vortex state (FC versus ZFC) and explains the appearance of hysteretic
effects, with results that are in good agreement with experiments. With the
new information at hand, the pinscape can be analyzed in much more detail via
deliberate state preparation `in between' the field- and zero-field-cooled
extremes.

We acknowledge financial support of the Fonds National Suisse through the NCCR
MaNEP. Research in Ames was supported by the U.S.\ DOE under contract
\#DE-AC02-07CH11358.


\begin{thebibliography}{99}
\bibitem{remark} Note that within the phenomenological approach, the Bean
critical state corresponds to a vanishing curvature $\alpha = 0$; the
diverging $\lambda_{\rm \scriptscriptstyle C} \to \infty$ can be attributed to
vortices penetrating deep into the sample.
\end{thebibliography}
\end{document}